\documentclass[12pt]{iopart}

\usepackage{bm}
\usepackage{graphicx}
\usepackage{times}
\usepackage{soul}
\usepackage{color}

\begin{document}

\title{Valley pseudospin relaxation of charged excitons in~monolayer MoTe$_2$}

\def \UW{Institute of Experimental Physics, Faculty of Physics, University of Warsaw, Pasteura 5, 02-093 Warsaw, Poland}

\def \LNCMI{Laboratoire National des Champs Magnetiques Intenses,
CNRS-UGA-UPS-INSA-EMFL, 25 rue des Martyrs, 38042 Grenoble, France}

\def \ETH{Institute for Quantum Electronics, ETH Zurich, CH-8093 Zurich, Switzerland}

\author{T. Smole\'nski\textsuperscript{1,2}, T. Kazimierczuk\textsuperscript{1}, M. Goryca\textsuperscript{1}, K. Nogajewski\textsuperscript{1}, M.~Potemski\textsuperscript{1,3}, P. Kossacki\textsuperscript{1}}

\address{$^1$ \UW}
\address{$^2$ \ETH}
\address{$^3$ \LNCMI}

\ead{Tomasz.Kazimierczuk@fuw.edu.pl}
\vspace{10pt}

\begin{abstract}
Zeeman effect induced by the magnetic field introduces a splitting between the two valleys at $K^+$ and $K^-$ points of the Brillouin zone in monolayer semiconducting transition metal dichalcogenides.
In consequence, the photoluminescence signal exhibits a~field dependent degree of circular polarization. We present a comprehensive study of this effect in the case of a~trion in~monolayer MoTe$_2$, 
showing that although time integrated data allows us to deduce a g-factor of the trion state, such an~analysis cannot be substantiated by the timescales revealed in the time-resolved experiments.
\end{abstract}

\section{Introduction}
Valley degree of freedom is one of the central points in the physics of two-dimensional crystals with honeycomb lattice, such as, for example graphene\cite{Novoselov_Science_2004} and ultrathin layers of transition metal dichalcogenides \cite{Mak_PhysRevLett_2010}.
This has been first theoretically recognized \cite{Rycerz_NatPhys_2007} and then largely studied in experiments, predominantly focused on the optical properties of monolayers of semiconductor transition metal dichalcogenides (S-TMDs).
Distinctly, the individual valleys at $K^+$ and $K^-$ points of the Brillouin zone in S-TMD monolayers can be accessed by using light with selected $\sigma^+$ or $\sigma^-$ circular polarization \cite{Cao_Nat.Commun._2012,Mak_NatNanotechnol_2012,Zeng_Nat.Nanotechnol._2012} 
Up to date, a wide variety of valley-related phenomena has been
demonstrated experimentally, including valley Zeeman effect, optical orientation of the valley pseudospin,
or valley coherence \cite{Aivazian_Nat.Phys._2015,Jones_Nat.Nanotechnol._2013,Smolenski_PRX_2016}. 

While most of the research has been so far focused on the neutral exciton (X) due its simple electronic structure, the valley degree of freedom as an integral 
element of all excitonic complexes, Is an interesting property to look over also in charged excitons (CXs). Similarly to the case of spin degree of freedom, 
different relaxation mechanisms for X and CX complexes may result in significantly different valley relaxation time --- 
a~quantity of paramount importance for any future applications. 

In this work we focus on the CX valley relaxation in monolayer MoTe$_2$. The benefit from choosing this particular is the long bright exciton PL
decay time in comparison with other monolayer semiconductor TMDs \cite{Robert_PhysRevB_2016_mote2}. Although the difference in the decay constants 
is not qualitative, a~longer decay time facilitates investigation of the valley effects.

The main motivation behind our study is to exploit the field-induced polarization of the negative trion in order to assess the value of the hole g-factor in monolayer MoTe$_2$.
The idea of such a~measurement is based on the notion that the difference in trion population between the two valleys is dictated by the Boltzmann factor, which in turn depends on the value of the Zeeman splitting. In our approach the valley polarization is studied by means of time-integrated photoluminescence spectroscopy in the magnetic field up to 10~T. Such an experiment is conducted for a range of excitation intensities, which allows us to address the issue of laser-induced heating of the sample. This kind of time-integrated investigation is additionally confronted with time-resolved data, in order to substantiate a~core assumption of thermal equilibrium between trions in the two valleys.

\section{Photoluminescence of monolayer {M{o}T{e}}$_2$}

\begin{figure}
\centering
\includegraphics{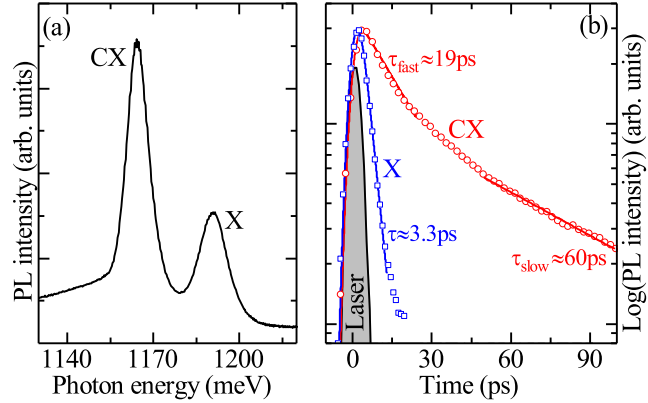}
\caption{(a) Representative PL spectrum of a MoTe$_2$ monolayer featuring both X and CX optical transitions. The spectrum was measured at $T=10$~K under pulsed, non-resonant excitation at $E\simeq1.7$~eV. (b) Time-resolved PL of X and CX. The solid lines represent mono-exponential fits to both the X decay profile and the fast and slow components of the CX decay profile. Each of the fitted profiles has been convolved with a Gaussian of standard deviation 2.2~ps corresponding to the overall temporal resolution of our experimental setup, as inferred from the width of the temporal profile of the backscattered laser light (shown for reference).\label{fig1:PL}}
\end{figure}

The experiments reported in this work were carried out on monolayer MoTe$_2$.
The flakes were mechanically exfoliated from bulk crystals with 2H structure purchased from HQ Graphene and deposited on a piece of chemically cleaned and oxygen-plasma ashed Si substrate covered with a 90-nm-thick SiO$_2$ layer. In order to obtain large, high quality monolayers, a two-stage, tape- and polydimethylsiloxane-based exfoliation technique was used. After the exfoliation, the flakes of interest were first identified by their characteristic optical contrast and then subjected to Raman scattering and atomic force microscopy measurements to unambiguously confirm their monolayer thickness.

The optical measurements were performed in a photoluminescence (PL) setup with a~spatial resolution
of about 1 $\mu$m. The sample was excited non-resonantly using a~femtosecond Ti:sapphire 
laser with repetition rate of 76~MHz. The PL signal was detected either with an InGaAs CCD camera or a streak camera with an S1
cathode for time-resolved measurements. The temperature of the sample was controlled using 
a~bath cryostat with a~variable-temperature insert. The cryostat was equipped with 
a~superconductive magnet producing magnetic field up to 10~T in the Faraday configuration. 

 \begin{figure*}
\centering
\includegraphics[width=0.9\textwidth]{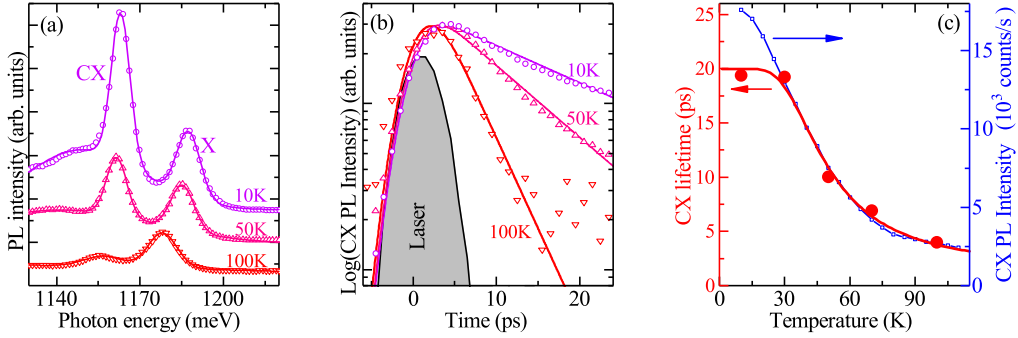}
\caption{(a) PL spectra of a MoTe$_2$ monolayer taken at different temperatures, but for fixed power and energy of the pulsed laser excitation, equal to, respectively, $0.72$~mW and $1.7$~eV. Solid lines represent multi-peak Gaussian fits to the measured spectra. (b) Temporal decay profiles of the CX PL acquired under the same conditions as the PL spectra in (a). Solid lines are Gaussian-convolved mono-exponential fits to the experimental data. The instrument response profile determined using the laser light backscattered from the sample surface is depicted for reference. (c) The lifetime (left axis) and PL intensity (right axis) of the CX optical transition plotted as a function of the temperature. Red solid curve represents the fit of the lifetime dependence with a simple theoretical model described in the text. 
\label{fig2:PL_vs_T}}
\end{figure*}

A representative PL spectrum of a monolayer MoTe$_2$ at $T=10$~K is presented in Fig. \ref{fig1:PL}(a).
In accordance with earlier studies \cite{Lezama_NanoLett_2015,Ruppert_NanoLett_2014}, the spectrum consists of two emission peaks, corresponding to the neutral exciton (X) and 
the charged exciton (CX). We corroborate this assignment with time-resolved measurements shown
in Fig. \ref{fig1:PL}(b). As in all monolayer S-TMDs, the neutral exciton
exhibits an ultra-fast decay on the order of single picoseconds, while the trion decays by an~order of magnitude slower\cite{Wang_ApplPhysLett_2015,Robert_PhysRevB_2016_lifetimes}. 
The experimentally measured lifetime of the neutral exciton of $\tau \approx 3.3$ ps only slightly exceeds the jitter of our setup ($\sigma = 2.2$ ps),
consistently with the measurements reported in Ref. \cite{Robert_PhysRevB_2016_mote2}. On the other hand, the decay of the charged exciton yields a~decay
constant of $\tau \approx 19$ ps, easily resolved by the streak camera. At longer times the CX decay profile deviates from the single exponential shape, possibly
due to population conversion from the dark states, but in further considerations we will only focus on the short-lived component.

\begin{figure}
\centering
\includegraphics{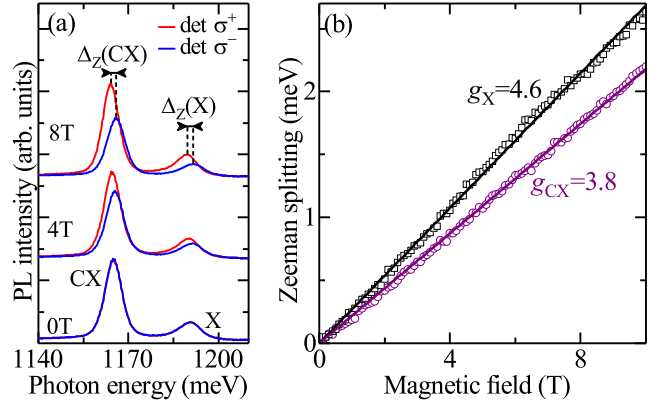}
\caption{(a) Circular-polarization-resolved PL spectra of a MoTe$_2$ monolayer measured at $T=10$~K under different magnetic fields applied perpendicularly to the monolayer plane. (b) Values of Zeeman splittings of the X and CX optical transitions determined as a function of magnetic field. Solid lines represent the linear fits to the X and CX data, which correspond to spectral $g$-factors of $g_\mathrm{X}=-4.6$ and $g_\mathrm{CX}=-3.8$, respectively.\label{fig3:magneto_PL}}
\end{figure}

The properties discussed above correspond to the case of low-temperature limit. Upon increasing the temperature, the spectrum undergoes two distinctive changes,
as shown in Fig. \ref{fig2:PL_vs_T}. First, both X and CX transitions are gradually red-shifted due to the temperature dependence of the MoTe$_2$ bandgap \cite{Lezama_NanoLett_2015}.
Secondly, the overall intensity of both emission lines diminishes. The quenching occurs at a~different rate for X and CX emission, which leads to temperature-induced changes in their relative
intensity. It is still under debate whether these changes reflect the mass action law \cite{Ross_NatCommun_2013} or rather the increasing role of phonon-trion scattering, acting as 
a~non-radiative decay channel for the trions \cite{Robert_PhysRevB_2016_lifetimes}. In order to shed more light on this issue we have performed a series of time-resolved measurements at different temperatures.
The results of three of such measurements are presented in Fig. \ref{fig2:PL_vs_T}(b). The data clearly shows an increase of the decay rate upon increasing the temperature. 
As shown in Fig. \ref{fig2:PL_vs_T}(c), with the exception of the lowest temperatures, the quenching of the PL intensity and the decrease of the CX decay time exhibit the same temperature dependence. The deviation for low temperatures might be related to the biexponential character of the decay profile, which is more pronounced in that temperature range.
The dependence presented in Fig. \ref{fig2:PL_vs_T}(c) can be fit by a~simple model of two competing decay channels: a temperature-independent radiative recombination and a thermally activated phonon-assisted decay:
\begin{equation}
\frac{1}{\tau} = \frac{1}{\tau_0} + \frac{1}{\tau_\mathrm{ph}} \exp\left(-\frac{E_\mathrm{ph}}{kT}\right).
\end{equation}
The solid line in Fig. \ref{fig2:PL_vs_T}(c) presents the result of such a fitting with $E_\mathrm{ph} = 14$~meV. We note that a similar dependence was observed in monolayer MoSe$_2$ in Ref. \cite{Robert_PhysRevB_2016_lifetimes} with $E^\mathrm{(MoSe_2)}_\mathrm{ph} = 33$~meV. We tentatively associate the difference between these two values with the difference in the $\mathrm{A_{1g}}$ phonon energies in the two materials, even though there is only partial quantitative agreement in this respect ($E^\mathrm{(MoSe_2)}_\mathrm{A_{1g}} = 30$~ meV vs $E^\mathrm{(MoTe_2)}_\mathrm{A_{1g}} = 21$~ meV \cite{Froehlicher_NanoLett_2015}) .

\section{Influence of the magnetic field on the charged exciton valley psuedospin relaxation \label{sec:timeintegrated}}

In order to study the valley-related phenomena we employed a circularly-polarized detection scheme in the PL experiment. Under such conditions we independently access excitons in the K$^+$ or K$^-$ valley by 
detecting, respectively, in the $\sigma+$ or $\sigma-$ polarization \cite{Mak_NatNanotechnol_2012}. In the absence of magnetic field two valleys are degenerate, yielding identical PL signal (see Fig. \ref{fig3:magneto_PL}(a)). Upon application of magnetic field in the Faraday geometry, the valley degeneracy is lifted, as evidenced by an increased difference between the PL spectra recorded in both polarizations.
The splitting of both the X and CX line grows linearly with magnetic field with a $g$ factor of $-4.6$ and $-3.8$ respectively. These values are consistent with earlier measurements carried out in high magnetic fields \cite{Arora_NanoLett_2016}.

\begin{figure*}[t]
\centering
\includegraphics[width=0.9\textwidth]{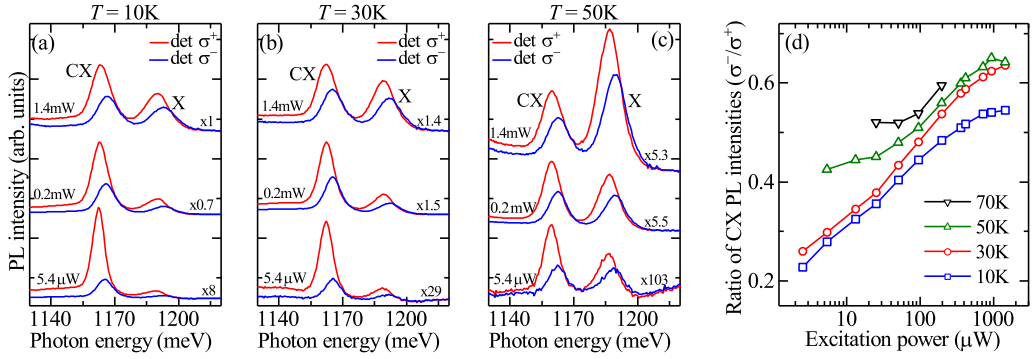}
\caption{(a-c) Circular-polarization-resolved PL spectra of a MoTe$_2$ monolayer measured under $B=10$~T at various (indicated) powers of pulsed-laser excitation and at different bath temperatures: 10~K~(a), 30~K~(b), and 50~K~(c). For the sake of clarity, the spectra acquired at different powers are vertically displaced. Moreover, they are also normalized by indicated factors, in order for the $\sigma^+$-polarized component of the CX transition to remain equally strong at each power/temperature.~(d) Exciation power dependence of the ratio of CX PL intensities measured in the $\sigma^-$ and $\sigma^+$ polarizations at various temperatures.
\label{fig4:magneto_PL_vs_T_and_P}}
\end{figure*}

As shown in Fig. \ref{fig3:magneto_PL}(a), the Zeeman splitting of the X and CX lines is associated with a difference in the overall PL intensity from the two valleys. The dominance of the lower
energy component clearly indicates that the effect is due to relaxation of the exciton populations from the K$^-$ to the K$^+$ valley. A systematic analysis of this effect is presented in Fig. \ref{fig4:magneto_PL_vs_T_and_P}.

A~comparison between pairs of PL spectra in each of Figs. \ref{fig4:magneto_PL_vs_T_and_P}(a-c) evidences that the degree of the relaxation is lower for higher excitation power used in the experiment. Similarly, a~comparison across the panels \ref{fig4:magneto_PL_vs_T_and_P}(a-c) reveals that the degree of the relaxation is lower for higher temperatures. The combined effect of these two factors is shown in Fig. \ref{fig4:magneto_PL_vs_T_and_P}(d). The simplest interpretation of the dependence on the excitation power is the notion that the actual temperature at the laser spot is increased by stronger excitation. In this view the degree of the relaxation at a given magnetic field is governed only by the temperature, but the temperature of the sample is accurately read only for vanishing excitation power.

For the sake of deeper analysis of the data in Fig. \ref{fig4:magneto_PL_vs_T_and_P}(d) we tentatively assume a~perfect thermal equilibrium between trion populations in the two valleys. The validity of this assumption will be challenged in the next section, but it allows us to link the experimentally measured ratio of the PL signals from both valleys to the intrinsic Zeeman splitting. In particular, the Boltzmann distribution dictates that
\begin{equation}
\frac{I_+}{I_-} = \exp\left( - \frac{g_\mathrm{CX,ini}\, \mu_\mathrm{B} B}{kT_\mathrm{eff}} \right), \label{eq:boltzmann}
\end{equation}
where $I_\pm$ denotes the population of trions in the K$^\pm$ valley, $T_\mathrm{eff}$ is the effective temperature at the measured spot, and $g_\mathrm{CX,ini}$ is a g-factor of the CX state. We note that $T_\mathrm{eff}$ is the temperature of the exciton system, which in the case of optical excitation can be higher than the lattice temperature.

We emphasize that the relevant g-factor of the CX state in Eq. \ref{eq:boltzmann} is different than the previously discussed \emph{spectral} g-factor $g_\mathrm{CX}\approx -3.8$, as the latter one is modified by the g-factor of the final state for the recombination.
The g-factor of the CX may be rather considered as a g-factor of a minority carrier (i.e., the hole in the case of a~negatively charged trion), since the two majority carriers form a singlet pair with no magnetic response.
The equation (\ref{eq:boltzmann}) can be transformed to extract the g-factor:
\begin{equation}
\frac{1}{g_\mathrm{CX,ini}} \frac{T_\mathrm{eff}}{T}  =  \frac{\mu_\mathrm{B} B}{kT} / \log \frac{I_-}{I_+}. \label{eq:dziwna_skala}
\end{equation}
The aim of introducing here the bath temperature $T$ is to make both sides of the equation dimensionless.

\begin{figure}
\centering
\includegraphics{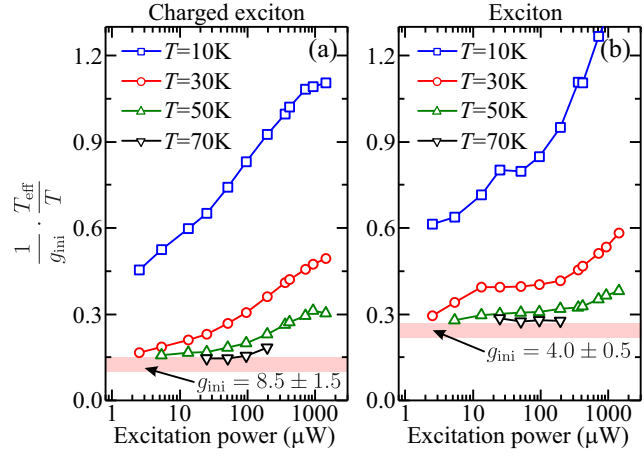}
\caption{(a-b) Pulsed-laser excitation power dependence of the ratio of effective valley pseudospin temperature $T_\mathrm{eff}$ and bath temperature $T$ divided by the initial state $g$-factor determined for CX (a) and X (b) optical transitions. The plotted quantity $(T_\mathrm{eff}/T)/g_\mathrm{ini}$ was obtained based on the ratio $R$ of the $\sigma^\pm$-polarized intensities of a given transition (measured at $B=10$~T at different temperatures $T$) using the following expression $\mu_BB/k_BT\ln(1/R)$, where $\mu_B$ stands for the Bohr magneton, while $k_B$ represents the Boltzmann constant. The light-red lines depicted in each panel correspond to the values of $(T_\mathrm{eff}/T)/g_\mathrm{ini}$ obtained at low excitation-power limit, which, for appropriately large bath temperatures (where $T_\mathrm{eff}\simeq T$), are determined by $1/g_\mathrm{ini}$, thus allowing to extract $g_\mathrm{ini}$ for both CX and X transitions, yielding $-8.5\pm1.5$  and $-4.0\pm0.5$, respectively.\label{fig5:magneto_PL_effective_T}}
\end{figure}

Figure \ref{fig5:magneto_PL_effective_T}(a) shows the results of application of the formula (\ref{eq:dziwna_skala}) to the data from Fig. \ref{fig4:magneto_PL_vs_T_and_P}(d). 
In such a presentation, each data series corresponding to different cryostat temperature tends to saturate at the same level for low excitation intensity. Such a behavior is expected, as for the lowest excitation powers the ${T_\mathrm{eff}}/{T}$ factor in Eq. \ref{eq:dziwna_skala} should converge to $1$. The remaining value should correspond to the inverse of the trion \emph{state} g-factor: ${1}/{g_\mathrm{CX,ini}}$. Based on our experimental data we estimate its value as $g_\mathrm{CX,ini} = -8.5 \pm 1.5$, which is marked in Fig. \ref{fig5:magneto_PL_effective_T}(a) with a light-red bar. The precision of our estimation is limited, as there is no clear-cut criterion for the saturation. Additionally, reaching the sufficiently low excitation power for the lowest bath temperature was not feasible due to progressively increasing accumulation time required in the PL experiment. On the other hand, the obtained value of $-8.5$ is consistent with the expectation for the valence band g-factor based on a simple addition of carrier spin, orbit, and valley contributions to the Zeeman effect \cite{Koperski_2DMat_2019}.  Moreover, by repeating the same procedure for the neutral exciton PL signal we obtain an estimation of $g_\mathrm{X,ini} = -4.0 \pm 0.5$ (see Fig. \ref{fig5:magneto_PL_effective_T}(b)), which is close to the actual neutral exciton g-factor of $-4.6$.

\section{Relaxation dynamics of the charged exciton valley pseudospin}

\begin{figure*}
\centering
\includegraphics[width=0.9\textwidth]{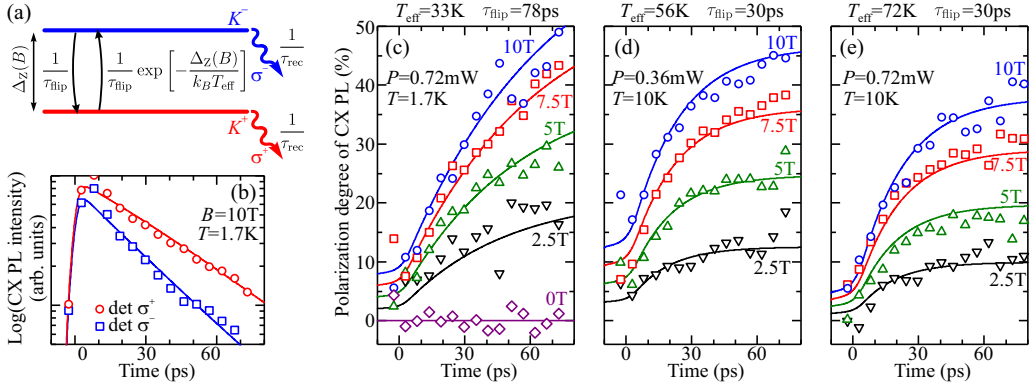}
\caption{(a) Scheme of transitions between the two Zeeman-split CX valley-pseudospin states, which were included in the rate equation model of the CX pseudospin relaxation. The Zeeman splitting $\Delta_\mathrm{Z}(B)$ is calculated using trion state g-factor of $8.5$. (b) Time-resolved CX PL decay profiles measured at $T=1.7$~K, $P=0.72$~mW, $B=10$~T in two different circular polarizations of detection (as indicated). (c-e) Temporal profiles of CX PL circular-polarization degree measured under different magnetic fields and various bath temperatures $T$ and/or excitation powers $P$: $T=1.7$~K, $P=0.72$~mW~(c), $T=10$~K, $P=0.36$~mW~(d), and $T=10$~K, $P=0.72$~mW~(e). Solid curves in panels (b-e) represent fits to the experimental data with the rate-equation model described in the text.\label{fig6:magneto_PL_time_res}}
\end{figure*}

A crucial element of the analysis presented in the previous section was a premise of thermal equilibrium between the CX populations in the two valleys. In order to verify this assumption we performed time-resolved PL measurements in two circular polarizations. At first sight, the results of such an experiment (see Fig. \ref{fig6:magneto_PL_time_res}(b)) resemble the decay shown earlier in Fig. \ref{fig1:PL}(b). However, a closer inspection reveals that the dynamics of the signal in the two circular polarization is not identical. In order to evidence these differences, instead of analyzing each polarization separately, we calculate at each point of time the instantaneous degree of polarization. 
Figures \ref{fig6:magneto_PL_time_res}(c-e) present such data series for three different experimental conditions.  

Our analysis of the measured transients is based on a rate-equation model presented schematically in Fig. \ref{fig6:magneto_PL_time_res}(a). The model assumes a single relaxation time $\tau_\mathrm{flip}$ for the valley pseudo-spin. The rate of inverse transition is dictated by the Boltzmann distribution with an effective temperature $T_\mathrm{eff}$. The resulting thermalization of the valley pseudo-spin is accompanied by radiative recombination from each valley. The initial populations of each valley are considered as free parameters to abstract from more complex processes occurring during an initial ultra-fast relaxation of the hot photocreated carriers. As shown with solid lines in Figs. \ref{fig6:magneto_PL_time_res}(c-e), we were able to fit each of the data series using a single effective temperature and relaxation time $\tau_\mathrm{flip}$. Their values are shown on top of each panel in Figs. \ref{fig6:magneto_PL_time_res}(c-e). A~comparison of panel (c) with (e) and panel (d) with (e) proves that both fitted parameters depend on, respectively, the bath temperature and the excitation power. 

Particularly important is the comparison between the fitted valley relaxation time and the previously measured CX decay time. Clearly, in all the cases shown in Figs. \ref{fig6:magneto_PL_time_res}(c-e) the $\tau_\mathrm{flip}$ is significantly longer than the low-temperature CX recombination time of $\tau_\mathrm{CX} \approx 20$~ps. It entails that the bulk of the PL is emitted \emph{before} the system reaches the equilibrium between the valleys. From this point of view, the lowering of the excitation power does not improve the situation,
giving no warranty that in the low power limit discussed in Section \ref{sec:timeintegrated} the polarization of the time-integrated PL signal is indeed governed by the g-factor of the initial state. Moreover, our data indicates that the disparity between the valley relaxation time and the CX recombination time is more severe at lower temperatures. It leads to a counter-intuitive conclusion that elevated temperatures can provide a better estimation of the initial state g-factor, despite a worse signal-to-noise ratio due to reduced difference between the PL signals in the two polarizations.

As evidenced by this discussion, the timescales revealed by the time-resolved measurements cannot explain why the data in Fig. \ref{fig5:magneto_PL_effective_T} converge at plausible values of the g-factors. This paradox can be perhaps explained by the variation of the initial polarization degree. As seen in Fig. \ref{fig6:magneto_PL_time_res}(c-e), at $t=0$ the valley polarization does not start at $0$, but at some finite field-dependent value. In our analysis we have treated these values as free parameters, as they originate from the interplay between little-known ultrafast cooling processes of the photo-excited carriers and the polarization of the residual carriers. It is possible that upon lowering the excitation power these starting values are getting closer to the saturation values, effectively reducing the impact of slow rate of consecutive valley pseudospin relaxation. Unfortunately, at this point we are unable to directly verify such a scenario due to limited sensitivity of the streak camera measurements.

\section{Summary}

In our work we have analyzed the CX valley polarization using both time-integrated and time-resolved detection of the PL signal. In either case we observed a strong dependence of the measured polarization on the bath temperature and the excitation power. The experimental data was described using simple models accounting for both these factors, which allowed us to extract parameters such as the g-factor of a CX state or a relevant valley pseudospin relaxation rate.

The result of the time-resolved experiment clearly shown that the premise of the equilibrium between the CX populations in two valleys is of no merit.
Yet, we show that analysis of the time-integrated data based on this assumption still leads to realistic value of the CX g-factor. We conclude that while such an approach inherently underestimates the Zeeman splitting, the introduced error can be mitigated by avoiding too low temperatures.

\section{Acknowledgments}
The work was supported by the ATOMOPTO project carried out within the TEAM programme of the Foundation for Polish Science co-financed by the European Union under the European Regional Development Fund, and National Science Center, Poland under project no. DEC-2015/17/B/ST3/01219.

\section{Bibliography}
\bibliography{manuscript}
\bibliographystyle{unsrt}

\end{document}